\documentclass[prb,showpacs,floatfix,twocolumn,superscriptaddress]{revtex4}
\usepackage{graphicx}
\usepackage{dcolumn}

\begin{document}

\newcommand*{\cm}{cm$^{-1}$\,}
\newcommand*{\mto}{MgTi$_2$O$_4$\,}
\newcommand*{\cis}{CuIr$_2$S$_4$\,}
%
\title{Optical study of MgTi$_2$O$_4$: Evidence for an orbital-Peierls state}
%
%
\author{J. Zhou}
\author{G. Li}
\author{J. L. Luo}
\author{Y. C. Ma}
\author{Dan Wu}
\affiliation{Beijing National Laboratory for Condensed Matter
Physics, Institute of Physics, Chinese Academy of Sciences,
Beijing 100080, P. R. China}
\author{B. P. Zhu}
\author{Z. Tang}
\author{J. Shi}
\affiliation{Department of Physics, Wuhan University, Wuhan, Hubei
430072, P.~R.~China}
\author{N. L. Wang}
\altaffiliation{Corresponding author}\email{nlwang@aphy.iphy.ac.cn}%
\affiliation{Beijing National Laboratory for Condensed Matter
Physics, Institute of Physics, Chinese Academy of Sciences,
Beijing 100080, P. R. China}
%
%
%

\begin{abstract}
Dimension reduction due to the orbital ordering has recently been
proposed to explain the exotic charge, magnetic and structural
transitions in some three-dimensional (3D) transitional metal
oxides. We present optical measurement on a spinel compound
MgTi$_2$O$_4$ which undergoes a sharp metal-insulator transition
at 240 K, and show that the spectral change across the transition
can be well understood from the proposed picture of 1D Peierls
transition driven by the ordering of $d_{yz}$ and $d_{zx}$
orbitals. We further elaborate that the orbital-driven instability
picture applies also very well to the optical data of another
spinel CuIr$_2$S$_4$ reported earlier.
\end{abstract}


\maketitle

%

\section{INTRODUCTION}

In low-dimensional electronic system, Fermi surface instability
often occurs at low temperature and drives the system into a
symmetry-breaking insulating state. However, such instability is
not expected to develop in a three-dimensional (3D) system.
Recently, two types of highly exceptional orderings were
discovered in two spinel compounds by Radaelli and co-workers:
octamer ordering in \cis\cite{Radaelli} and helical (or chiral)
ordering in \mto\cite{Schmidt}. In both cases, sharp
metal-insulator transitions (MIT) and spin-dimerizations
associated with the structural distortions occur
simultaneously\cite{Furubayashi,Isobe}. Those extraordinary
magnetic, charge and structural transitions have generated a great
deal of interest and
discussion\cite{Cao,Croft,Popovic,Wang,Matteo,Zhou,Khomskii,Radaelli2}.
Recently, Khomskii and Mizokawa\cite{Khomskii} suggested that the
orbital degree of freedom plays a key role in such transition: the
ordering of the orbitals makes the electrons travel exclusively
along certain chains which effectively leads to the reduction of
the dimensionality from 3D to 1D. Then, the very strange octamer
and chiral structural change can be easily understood from the
so-called orbitally-driven Peierls state involving the formation
of a dimerized state with alternating strong and weak bonds along
the chains formed by Ir or Ti ions. The losing of 3D nature driven
by the orbital ordering has recently been generalized to other
transition metal oxides, for example in 3D pyrochlore
Tl$_2$Ru$_2$O$_7$ where possible formation of 1D Haldane chains
was indicated\cite{Lee}. It is further believed that such kind of
dimension reduction opens up a new direction in research into
transitional metal oxides\cite{Brink}.

Spinels have a general formula AB$_2$X$_4$, where A and B are
metallic ions and X is an anion, such as O or S. For many spinels,
including \cis and \mto, A-site ions have fully filled energy
levels, while B-site ions, locating at the center of X$_6$
octahedra, have partially-filled d levels, which thus determine
the low-lying excitations. Those B-site ions (or BX$_6$ octahedra)
are arranged in chains through corner-sharing tetrahedra (i.e.
pyrochlore lattice). Although \cis and \mto offer good
opportunities for investigating possible novel type of Peierls
transition in a 3D system, few experiments were performed on those
systems due to the difficulty of synthesizing high quality samples
or single crystals. This is the case particularly for \mto owing
to the low oxidation state of Ti$^{3+}$ ions. We recently prepared
single phase polycrystalline samples of Mg$_{1+x}$Ti$_{2-x}$O$_4$
using a novel Plasma arc melting method. The samples obtained by
this method are extremely dense. A very shinny and metallic bright
surface is obtained after fine polishing. Here we report the
optical study on \mto compounds. In combination with earlier data
collected on \cis, we show that the orbital-Peierls transition
picture provides excellent explanation for the spectral change
across the MIT for both compounds. Therefore, the study provides
strong support for the picture of orbital-driven 1D physics in
such 3D systems.

\section{EXPERIMENTAL RESULTS}

Fig. 1 displays the T-dependent dc resistivity and specific heat
curves measured by a Quantum Design PPMS. The detailed description
about sample preparation and other characterizations were
presented elsewhere\cite{Zhu}. The $\rho(T)$ increases sharply at
T$_{MIT}$=240 K. A small hysteresis could be seen for the
$\rho(T)$ curve. Accompanying the resistivity change, the specific
heat curve displays a peak at the same temperature region. The
hysteresis in $\rho(T)$ and the peak anomaly rather than a
$\lambda$-shape in specific heat imply that the transition is of
the first-order in nature. We note that the T$_{MIT}$ is lower
than the reported values in literature, and the peak in specific
heat is somewhat broad. This could be ascribed to the difference
in the amount of Ti ions being substituted by Mg ions.
Additionally, the $\rho(T)$ curve above T$_{MIT}$ also has a
negative slope, suggesting non-metallic nature even in the high-T
phase. This is similar to \cis where a negative slope (though less
value) is also seen even in single crystal sample\cite{Wang}. We
think that this behavior may arise from two effects: one is the
intrinsic "bad-metal" nature seen quite often in transitional
metal oxides for which the mean free path of electrons might be
close to the lattice constant. The other is the partial
substitutions of Ti by Mg ions. The randomly distributed
substitutions make the electrons further localized, and also
broadens the transitions as seen in resistivity and specific heat
curves. However, for consistency, we still use the term "MIT" in
the rest part of the paper.

\begin{figure}[t]
\centerline{\includegraphics[width=2.3in]{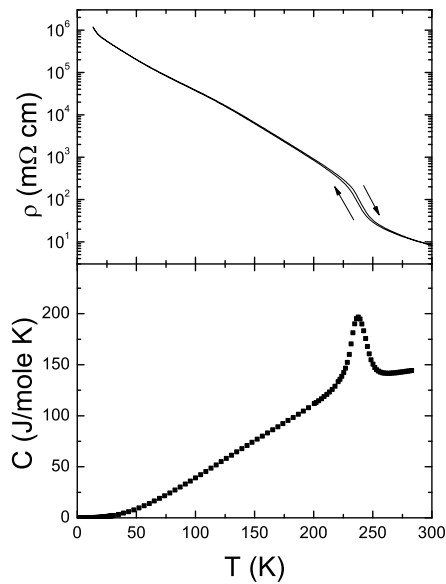}}%
\vspace*{-0.2cm}%
\caption{The dc resistivity and specific heat
\textit{vs}. temperature for \mto.}%
\label{1}
\end{figure}
\begin{figure}[t]
\centerline{\includegraphics[width=3.1in]{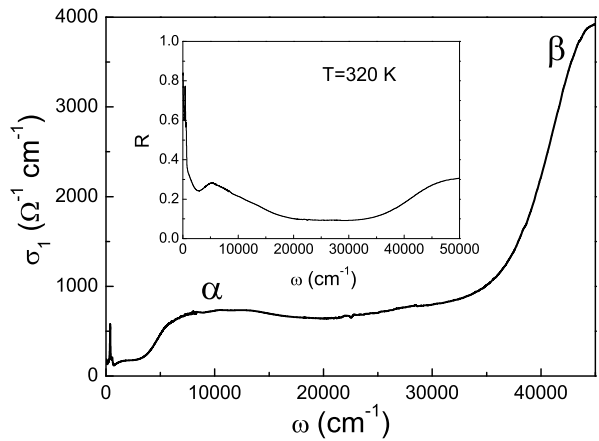}}%
\vspace*{-0.2cm}%
\caption{Optical reflectance and conductivity spectra for \mto at 320 K}%
\label{fig2}
\end{figure}

The frequency-dependent reflectance $R(\omega)$ was measured from
40 $cm^{-1}$ to 50,000 $cm^{-1}$ at different T on a Bruker 66v/s
and a grating type spectrometer, respectively, using an
\textit{in-situ} gold (below 15,000 cm$^{-1}$) and aluminum (above
10,000 cm$^{-1}$) overcoating technique. Since the material has a
3D cubic structure at high T, we can determine its optical
constants from the reflectance measurement on such high-dense
polycrystalline sample. Fig. 2 shows the optical reflectance and
conductivity spectra at 320 K over broad frequencies. The spectra
show two apparent interband transitions. A weak interband
transition, labelled as $\alpha$, starting from about 4,000 \cm
(0.5 eV) to 20,000 \cm (2.5 eV) is due to the transition from
$t_{2g}$ to $e_g$ bands, the strong one, $\beta$ peak, with onset
near 36,000 \cm (4.5 eV) is due to the transition from O$_{2p}$
bands to unoccupied states of $t_{2g}$ bands, as we shall explain
in detail below. The low-$\omega$ reflectance increases towards
unity, evidencing the conducting carrier response. However, the
reflectance values are still very low, so that the low-$\omega$
conductivity extracted is almost flat, without showing Drude-like
peak. The two sharp peaks below 700 \cm are infrared phonons.

\begin{figure}[t]
\centerline{\includegraphics[width=3.0in]{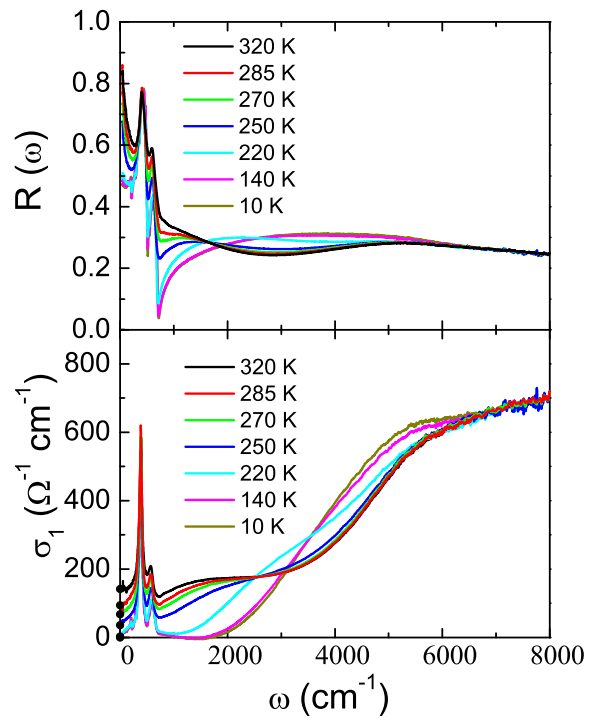}}%
\vspace*{-0.2cm}%
\caption{(Color online) The temperature dependence of the
reflectance and conductivity spectra in the frequency range of
0--8000 \cm. The black dots are corresponding dc conductivity
values.
The inset is an expanded plot of the low frequency region.}%
\label{fig3}
\end{figure}

Fig. 3 shows the R($\omega$) and $\sigma_1(\omega$) spectra at
different temperatures below 8,000 \cm (1 eV). As T decreases, the
low-$\omega$ R($\omega$) decreases, meanwhile the R($\omega$)
between 1,600 and 6,000 \cm increases. As a result, the broad
minimum centered near 3,000 \cm at high T gradually disappears and
the R($\omega$) displays a single broad peak above the phonon
peaks. In $\sigma_1(\omega$) spectra, the value below the onset of
interband transition $\sim$4,000 \cm is very low, without showing
Drude-like peak. In fact, the conductivity tends to decrease with
decreasing $\omega$. The higher values between 200 and 600 \cm are
due to phonons, which should be subtracted when trying to isolate
the electronic contribution. Those results suggest that the
electrons are rather localized even above T$_{MIT}$. In addition,
we find that the low-$\omega$ conductivity decreases with
decreasing T, furthering evidencing the non-metallic T-dependence.
Nevertheless, there is no gap in $\sigma_1(\omega$) at high T
(above MIT), the extrapolated values at zero frequency are in good
agreement with the dc resistivity data. Another remarkable feature
is that the optical spectra show very dramatic change as the
temperature decreases across the MIT. It is noted that the
R($\omega$) at our lowest measurement frequency decreases sharply
from about 0.7 at 250 K to about 0.5 at 220 K, and the sharp
increasing feature towards unity at zero frequency for T$>$250 K
vanishes completely. Such low-$\omega$ R($\omega$) is a
characteristic response behavior of an insulator. Correspondingly,
we find a rather rapid removal of low-$\omega$ spectral weight in
$\sigma_1(\omega$), leading to the formation of an energy gap. It
is noteworthy that the removal of spectral weight shifts only to
the region between 3,000 and 6,000 \cm, as a result, the onset of
interband transition occurs at a reduced energy. Above 6,000 \cm,
the optical spectra are almost T-independent. From
$\sigma_1(\omega$) at the lowest measurement T=10 K, it is easy to
identify the optical gap being about 2,000 \cm (0.25 eV).
Additionally, there appears an abrupt change in phonon structure
for temperature below and above T$_{MIT}$. In Fig. 4, we show the
$\sigma_1(\omega$) spectra in an expanded plot of the low
frequency region. As seen clearly, a number of new phonon modes
appears and the splitting of phonon mode occurs just below the
transition. Those results are consistent with the first-order
structural change with a lowering of lattice symmetry. Phonon
modes in this material were analyzed in detail
previously,\cite{Popovic} in this work we shall limit our
discussion to the electronic behavior.

\begin{figure}[t]
\centerline{\includegraphics[width=3.1in]{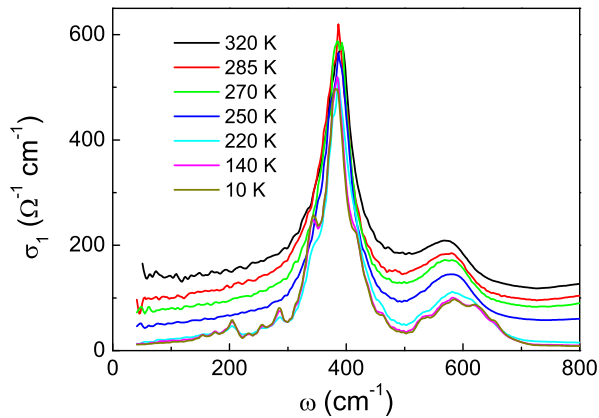}}%
\vspace*{-0.2cm}%
\caption{(Color online) An expanded plot of the low-frequency optical conductivity spectra at different temperatures.}%
\label{fig4}
\end{figure}

\section{DISCUSSIONS}

Let us now analyze the spectral change across the MIT. In the
spinel structure, the Ti$^{3+}$ ion locates at the center of
TiO$_6$ octahedron. As we mentioned above, those octahedra are
arranged in 1D chains (along six different directions) through
corner-sharing tetrahedra. At high T, the Ti 3d levels are split
into triply degenerate $t_{2g}$ level (orbitals $d_{xy}$,
$d_{zx}$, and $d_{yz}$) and a doubly degenerate $e_g$ level
(orbitals $d_{z^2-r^2}$ and $d_{x^2-y^2}$) under local cubic
environment. Since Ti$^{3+}$ has a 3d$^1$ electron configuration,
this single electron occupies the $t_{2g}$ level with equal
distribution on the $d_{xy}$, $d_{yz}$, $d_{zx}$ orbitals, leading
to the partially filled $t_{2g}$ band. The $e_g$ orbitals are
completely empty. The O 2\textit{p} levels are fully occupied, and
locate far away from the Fermi level, although some hybridizations
with Ti 3\textit{d} orbitals exist. This simple picture for the
band structure is supported by the first-principle
calculation\cite{Schmidt}. The optical spectra can be easily
understood from this simple picture, as illustrated in Fig 4. The
lowest \textit{interband} transition is from the occupied states
of $t_{2g}$ bands to empty $e_g$ states. This corresponds to the
$\alpha$ peak. Because the direct \textit{d-d} transition is
forbidden, the observed weak transition is largely due to the
hybridization with O 2\textit{p} band which thus makes the
transition allowable. The transition from occupied O 2\textit{p}
band to unfilled part of $t_{2g}$ band corresponds to the $\beta$
peak. However, the observed interband transition energies are
lower than the values obtained from the band structure
calculation\cite{Schmidt}. As the $t_{2g}$ bands are partially
filled, the \textit{intraband} transition should result in
metallic response at low frequency. Experimentally, the missing of
Drude-like peak suggests that the carriers are rather localized.
As discussed above, we think that this could be due to a
combination of bad-metal nature of the material and the disorder
effect caused by the substitution of Mg for Ti sites. We believe
that a successful growth of single crystal with better chemical
stoichiometry would reduce such non-metallic response.
Nevertheless, the entire physical picture would not be affected.

\begin{figure}[t]
\centerline{\includegraphics[width=3.0in]{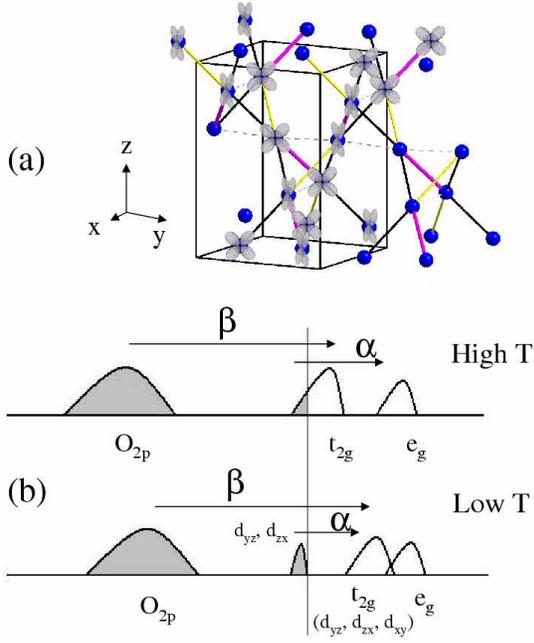}}%
\vspace*{-0.2cm}%
\caption{(Color online) (a) Orbital ordering of \mto at low T. The
B-site ions (blue dots) are arranged in chains through
corner-sharing tetrahedra. The short (pink), intermediate (black)
and long (yellow) bonds together with $d_{yz}$ and $d_{zx}$
orbitals along the chains (four directions [0,1,1], [0,1,-1],
[1,0,1], and [1,0,-1]) are displayed. (b) Schematic diagram of the
electronic states above and below the MIT temperature. At high T,
the 3d$^1$ electron of Ti$^{3+}$ ion occupies the $t_{2g}$ bands
with equal distribution on the $d_{xy}$, $d_{yz}$ and $d_{zx}$
orbitals. The $e_g$ bands are at higher energy. The two arrows
indicates the two interband transitions. Below $T_{MIT}$, the Ti
3d$^1$ electron occupies the $d_{yz}$ and $d_{zx}$ orbitals,
leading to two quarterly filled 1D $d_{yz}$ and $d_{zx}$ bands,
which were further split into lower and upper sub-bands due to
Peierls instability. Then the lower sub-bands are fully occupied,
the upper sub-bands, being mixed
with $d_{xy}$ band, are empty, and a gap forms between them.}%
\label{fig5}
\end{figure}

Since an orbital has a specific shape, an important characteristic
for spinels is that the spacial orientations of $d_{xy}$, $d_{yz}$
and $d_{zx}$ orbitals are along the 1D chain directions forms by
the Ti site ions, respectively. As a result, the bands formed by
those orbital levels have 1D characteristic and are susceptible to
Peierls instability. This is believed to be the origin of the
first order structural change and accompanied MIT. At low T phase,
the lattice symmetry is lowered to tetragonal\cite{Schmidt}.
According to the band structure
calculations,\cite{Khomskii,Schmidt} this tetragonal distortion
increases the bandwidths of the $d_{zx}$ and $d_{yz}$ bands, and
decreases that of the $d_{xy}$. With one electron per Ti ion, the
electron occupies the lowest doubly degenerate {$d_{zx}$ and
$d_{yz}$ level. The $d_{xy}$ level is pushed to higher energy and
is unoccupied. In this case, the orbital in $d_{yz}$ (or $d_{zx}$)
chain is quarterly occupied (i.e. the orbital is occupied in every
two sites), which makes the {$d_{zx}$ (or $d_{yz}$) band quarterly
filled. This leads to the Pererls instability and to the formation
of a tetramerization superstructure in the zx and yz directions:
an ordered arrangement of short, intermediate and long bonds along
$d_{yz}$ and $d_{zx}$ chains (four directions [0,1,1], [0,1,-1],
[1,0,1], and [1,0,-1]), respectively, as illustrated in the upper
panel of Fig. 4.\cite{Khomskii} Associated with the instability,
the $d_{zx}$ and $d_{yz}$ bands are split, respectively; the lower
parts of the split $d_{zx}$ and $d_{yz}$ bands are fully occupied,
the upper part, together with the $d_{xy}$ band, are separated by
a gap from the lower part of $d_{zx}$ and $d_{yz}$ bands.
Obviously, the observed gap in optical conductivity below
$T_{MIT}$ is due to the \textit{interband} transition from
occupied $d_{zx}$ and $d_{yz}$ bands to the unoccupied part of the
$t_{2g}$ manifold (upper part of $d_{zx}$, $d_{yz}$ mixing with
$d_{xy}$ band). The gap value of 2$\Delta\approx$2000 \cm leads to
2$\Delta/k_BT_{MIT}\approx$12. Considering that many systems with
charge-density wave transitions have $\Delta\gg$$k_BT_{MIT}$, the
gap value in \mto is not surprisingly large. Because the observed
$\alpha$ transition has a single broad peak with onset at lower
energy in comparison with the situation at high T, we believe that
the unoccupied $t_{2g}$ part has some overlap with $e_g$ bands.

\begin{figure}[t]
\centerline{\includegraphics[width=2.8in]{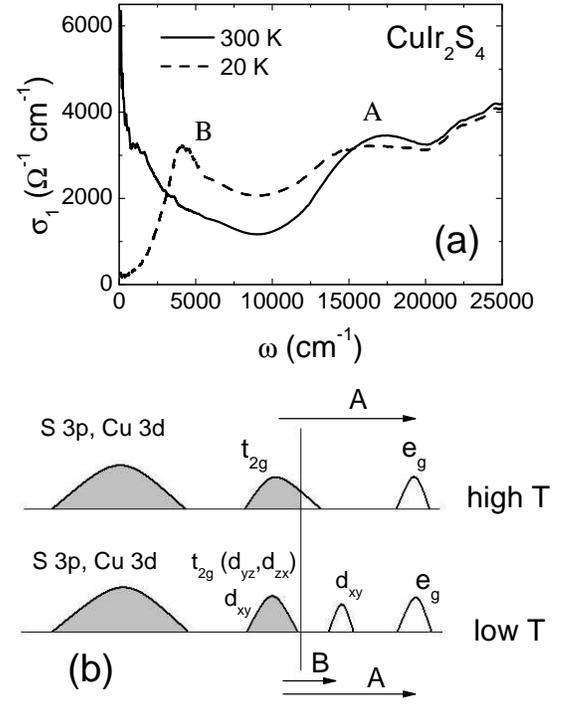}}%
\vspace*{-0.2cm}%
\caption{(a) The optical conductivity of \cis at 300 K and 20 K
taken from ref. [8]. (b) Schematic diagram of the electronic
states above and below the MIT temperature. At high temperature,
the 5.5 d-electron of Ir$^{3.5+}$ ion occupies the $t_{2g}$ bands
with equal probability on the $d_{xy}$, $d_{yz}$ and $d_{zx}$
orbitals. Arrow A indicates an interband transition from occupied
$t_{2g}$ to empty $e_g$ states. Below $T_{MIT}$, the structure
distortion splits the $t_{2g}$ bands. The doubly degenerate
$d_{yz}$ and $d_{zx}$ orbitals are fully occupied, however, the 1D
band along the $d_{xy}$ chain becomes quarterly filled, which was
further split due to Peierls instability. As a result, new states
above the Fermi level (i.e. upper sub-band of the split $d_{xy}$
band) were created. Arrow B indicates a new interband transition
from occupied $t_{2g}$ to
those new states.}%
\label{fig6}
\end{figure}

It deserves to remark that the above orbital-ordering picture
applies also extremely well to the optical data of \cis, which has
5.5 electrons in $t_{2g}$ bands. Optical spectra of \cis were
reported by us\cite{Wang} prior to the proposal of orbital-driven
instability picture, so the data were not analyzed in terms of the
orbital-Peierls transition scenario previously. To help readers
easily follow our discussion, we reproduce two conductivity curves
at 300 K and 20 K, being above and below T$_{MIT}$, as shown in
Fig. 6(a). Above T$_{MIT}$, the $\sigma_1(\omega$) displays a
Drude-like component at low frequency and an interband transition
peak (labelled as A) near 2 eV. Since at high T, every Ir ion is
equivalent, the 5.5 electrons have equal probability in
distribution among three degenerate $d_{xy}$, $d_{yz}$, $d_{zx}$
orbitals of $t_{2g}$ manifold, leading to the partially filled
$t_{2g}$ bands and completely empty $e_g$ bands. It is easy to
understand that the Drude component originates from the
\textit{intraband} transition of three degenerate $d_{xy}$,
$d_{yz}$, $d_{zx}$ bands (or $t_{2g}$ bands), while the 2 eV peak
originates from the \textit{interband} transition from occupied
$t_{2g}$ to empty $e_g$ bands. Such transitions are schematically
shown in the picture of Fig. 6 (b). However, below T$_{MIT}$, a
gap opens in $\sigma_1(\omega$) together with a new peak (labelled
as B) developing near 0.5 eV. Structurally, \cis undergoes a
octamer ordering with inequivalent Ir$^{3+}$ and Ir$^{4+}$ ions.
Ir$^{3+}$ has $t_{2g}^6$ configuration, Ir$^{4+}$ has $t_{2g}^5$
configuration. The octahedral distortion splits the $t_{2g}$
levels and lifts up the $d_{xy}$. Consequently, the doubly
degenerate $d_{zx}$ and $d_{yz}$ orbitals are fully filled, only
the $d_{xy}$ orbital of Ir$^{+4}$ ion is left to be partially
occupied. Then, one can identify 3/4-filled chains along $d_{xy}$
orbital directions, which also leads to a Pererls instability and
formation of a tetramerization superstructure in $d_{xy}$ orbital
directions with an ordered arrangement of short, intermediate and
long bonds as that in \mto (Fig. 2 of ref. [11]). This Peierls
instability splits the $d_{xy}$ band into two subbands, which
therefore creates new states above the Fermi level (upper sub-band
of the split $d_{xy}$ band). The interband transition from
occupied $t_{2g}$ to those states results in the new excitation
peak B as observed in optical conductivity.

The above optical study leads to tentative identification of the
novel orbital Peierls state in such 3D spinel compounds. It
highlights the 1D physics driven by the ordering of the orbital
degree of freedom in those compounds, despite of their 3D
structure. Recently, the orbital ordering picture was also
employed to other systems with similar phenomena, like
NaTiSi$_2$O$_8$\cite{Isobe2,Wezel,Streltsov},
La$_4$Ru$_2$O$_{10}$\cite{Khalifah,Eyert,Wu},
VO$_2$\cite{Haverkort}, and many other transition metal oxide
compounds\cite{Khomskii2,Efremov}. It would be very interesting to
study or reexamine their optical and other physical properties of
in terms of orbital-ordering picture.

\section{CONCLUSIONS}

The optical study on \mto indicates that the orbital degree plays
a crucial role in explaining the spectral evolution with T.
Although the $t_{2g}$ spinel systems have a 3D structure, they
behave more 1D-like. At high T, \mto has a cubic structure, the
low-lying excitations are dominated by the three degenerate 1D
bands formed by the chains of Ti-site ions through corner-sharing
tetrahedra. While at low T, the Ti 3d$^1$ electron tends to occupy
the doubly degenerate $d_{yz}$ and $d_{zx}$ orbitals, leading to
two quarter-filled 1D $d_{yz}$ and $d_{zx}$ bands. The optical
transitions could be well explained by the Peierls splitting of
the two 1D bands. We further elaborate that the orbital-ordering
picture applies also extremely well to the optical data of \cis.
The application of the 1D physics to 3D compounds with specific
orbital occupations has important implication for understanding
similar phenomena in other transitional metal oxides.

Acknowledgements: This work is supported by National Science
Foundation of China, the Knowledge Innovation Project of Chinese
Academy of Sciences, and the Ministry of Science and Technology of
China (973 project No. 2006CB601002).

%
%


\begin{thebibliography}{99}
\bibitem{Radaelli} P. G. Radaelli, Y. Horibe, M. J. Gutmann,
H. Ishibashi, C. H. Chen, R. M. Ibberson, Y. Koyama, Y.-S. Hor, V.
Kiryukhin, and S.-W. Cheong, Nature (London) \textbf{416}, 155
(2001).

\bibitem{Schmidt} M. Schmidt, W. Ratcliff II, P. G. Radaelli, K. Refson,
N. M. Harrison, and S. W. Cheong, Phys. Rev. Lett. \textbf{92},
056402 (2004).

\bibitem{Furubayashi} T. Furubayashi, T. Matsumoto, T. Hagino, and S.
Nagata, J. Phys. Soc. Jpn. \textbf{63}, 3333 (1994).

\bibitem{Isobe} M. Isobe and Y. Ueda, J. Phys. Soc. Jpn. \textbf{71}, 1848 (2002).

\bibitem{Cao} G. Cao, T. Furubayashi, H. Suzuki, H. Kitazawa, T.
Matsumoto, Y. Uwatoko, Phys. Rev. B \textbf{64}, 214514 (2001).

\bibitem{Croft} M. Croft, W. Caliebe, H. Woo, T. A. Tyson, D. sills, Y. S. Hor, S-W. Cheong,
V. Kiryukhin, and S. J. Oh, Phys. Rev. B \textbf{67}, 201102(R)
(2003).

\bibitem{Popovic} Z. V. Popovic, G. De Marzi, M. J. Konstantinovic,
A. Cantarero, Z. Dohcevic-Mitrovic, M. Isobe, and Y. Ueda, Phys.
Rev. B \textbf{68}, 224302 (2003).

\bibitem{Wang} N. L. Wang, G. H. Cao, P. Zheng, G. Li, Z. Fang, T. Xiang,
H. Kitazawa, and T. Matsumoto, Phys. Rev. B \textbf{69}, 153104
(2004).

\bibitem{Matteo} S. Di Matteo, G. Jackeli, C. Lacroix, and N. B. Perkins, Phys. Rev. Lett. \textbf{93}, 077208 (2004).

\bibitem{Zhou} H. D. Zhou and J. B. Goodenough, Phys. Rev. B \textbf{72}, 045118 (2005).

\bibitem{Khomskii} D. I. Khomskii and T. Mizokawa, Phys. Rev. Lett. \textbf{94}, 156402 (2005).

\bibitem{Radaelli2} P. G. Radaelli, New J. Phys. \textbf{7}, 53
(2005).

\bibitem{Lee} S. Lee, J.-G. Park, D. T. Adroja, D. Khomskii, S. Streltsov, K. A. Mcewen,
H. Sakai, K. Yoshimura, V. I. Anisimov, D. Mori, R. Kanno and R.
Ibberson, Nature Mater. \textbf{5}, 471 (2006).

\bibitem{Brink} J. van den Brink, Nature Mater. \textbf{5}, 427 (2006).

\bibitem{Zhu} B. P. Zhu, Z. Tang, L. H. Zhao, L. L. Wang, C. Z. Li, D. Yin, Z. x. Yu, W. F. Tang,
R. Xiong, J. Shi, and X. F. Ruan, Materials Letters (in press).

\bibitem{Isobe2} M. Isobe, E. Ninomiya, A. N. Vasilev, and Y. Ueda, J. Phys. Soc. Jpn. \textbf{71}, 1423 (2002).

\bibitem{Wezel} J. van Wezel and J. van den Brink, cond-mat/0512591.

\bibitem{Streltsov} S. V. Streltsov, O. A. Popova, and D. I. Khomskii, Phys. Rev. Lett. 96, 249701 (2006).

\bibitem{Khalifah} P. Khalifah, R. Osborn, Q. Huang, H. W. Zandbergen, R. Jin, Y. Liu,
D. Mandrus, and R. J. Cava, Science \textbf{297}, 2237 (2002).

\bibitem{Eyert} V. Eyert, S. G. Ebbinghaus, and T. Kopp, Phys. Rev. Lett. 96, 256401 (2006).

\bibitem{Wu} H. Wu, Z. Hu, T. Burnus, J. D. Denlinger, P. G. Khalifah, D. G. Mandrus, L.-Y. Jang,
H. H. Hsieh, A. Tanaka, K. S. Liang, J. W. Allen, R. J. Cava, D.
I. Khomskii, and L. H. Tjeng, Phys. Rev. Lett. 96, 256402 (2006).

\bibitem{Haverkort} M. W. Haverkort, Z. Hu, A. Tanaka, W.
Reichelt, S. V. Streltsov, M. A. Korotin, V. I. Anisimov, H. H.
Hsieh, H. J. Lin, C. T. Chen, D. I. Khomskii, and L. H. Tjeng,
Phys. Rev. Lett. \textbf{95}, 196404 (2005).

\bibitem{Khomskii2} D. I. Khomskii, Phys. Scr. \textbf{72},
CC8 (2005).

\bibitem{Efremov} M. Efremov, J. van den Brink, and D. I. Khomskii, Nature Materials. \textbf{3}, 853
(2004).

\end{thebibliography}
\end{document}